\documentclass{article}

\usepackage{arxiv}

\usepackage[utf8]{inputenc} 
\usepackage[T1]{fontenc}    
\usepackage{hyperref}       
\usepackage{url}            
\usepackage{booktabs}       
\usepackage{amsfonts}       
\usepackage{nicefrac}       
\usepackage{microtype}      
\usepackage{lipsum}		
\usepackage{graphicx}
\usepackage{natbib}
\usepackage{doi}
\usepackage{graphicx}
\usepackage{amsmath}
\usepackage{amssymb}
\usepackage{amsfonts}
\usepackage{braket}
\usepackage{float}
\usepackage[table,xcdraw]{xcolor}
\usepackage{cite}
\usepackage{subfig}

\usepackage{setspace}
\usepackage{multirow}
\usepackage[english, ruled, linesnumbered]{algorithm2e}
\usepackage{tikz}
\usepackage{url}

\newcommand\copyrighttext{%
\textcopyright 2021 IEEE. Personal use of this material is permitted. Permission from IEEE must be obtained for all other uses, in any current or future media, including reprinting/republishing this material for advertising or promotional purposes, creating new collective works, for resale or redistribution to servers or lists, or reuse of any copyrighted component of this work in other works. \url{https://doi.org/10.1109/TC.2021.3051559}.
}

\newcommand\copyrightnotice{%
\begin{tikzpicture}[remember picture, overlay, red]
\node[anchor=north, yshift=-20pt] at (current page.north) {\fbox{\parbox{\dimexpr\textwidth-\fboxsep-\fboxrule\relax}{\copyrighttext}}};
\end{tikzpicture}%
}

\title{Classical Artificial Neural Network Training Using Quantum Walks as a Search Procedure}

\author{ \href{https://orcid.org/0000-0002-6527-7065}{\includegraphics[scale=0.06]{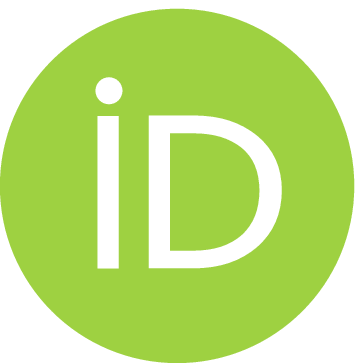}\hspace{1mm}Luciano S. de Souza}\thanks{95, R. Manuel de Medeiros, 35 - Dois Irmãos, Recife - PE.} \\
	Departamento de Estat\'{i}stica e Inform\'{a}tica\\
	Universidade Federal Rural de Pernambuco\\
	Recife, Brasil \\
	\texttt{lucianoserafimdesouza@gmail.com} \\
	\And
	\href{https://orcid.org/0000-0002-2672-7801}{\includegraphics[scale=0.06]{orcid.eps}\hspace{1mm}Jonathan H. A. de Carvalho} \\
	Centro de Inform\'{a}tica\\
	Universidade Federal de Pernambuco\\
	Recife, Brasil \\
	\texttt{hcarvalho.jon@gmail.com} \\
	\And
	\href{https://orcid.org/0000-0002-2131-9825}{\includegraphics[scale=0.06]{orcid.eps}\hspace{1mm}Tiago A. E. Ferreira} \\
	Departamento de Estat\'{i}stica e Inform\'{a}tica\\
	Universidade Federal Rural de Pernambuco\\
	Recife, Brasil \\
	\texttt{tiago.espinola@ufrpe.br} \\
}

\renewcommand{\shorttitle}{Classical Artificial Neural Network Training Using Quantum Walks as a Search Procedure}

\hypersetup{
pdftitle={A template for the arxiv style},
pdfsubject={q-bio.NC, q-bio.QM},
pdfauthor={David S.~Hippocampus, Elias D.~Striatum},
pdfkeywords={First keyword, Second keyword, More},
}

\begin{document}

\maketitle

\copyrightnotice

\begin{abstract}

This paper proposes a computational procedure that applies a quantum algorithm to train classical artificial neural networks. The goal of the procedure is to apply quantum walk as a search algorithm in a complete graph to find all synaptic weights of a classical artificial neural network. Each vertex of this complete graph represents a possible synaptic weight set in the $w$-dimensional search space, where $w$ is the number of weights of the neural network. To know the number of iterations required \textit{a priori} to obtain the solutions is one of the main advantages of the procedure. Another advantage is that the proposed method does not stagnate in local minimums. Thus, it is possible to use the quantum walk search procedure as an alternative to the backpropagation algorithm. The proposed method was employed for a $XOR$ problem to prove the proposed concept. To solve this problem, the proposed method trained a classical artificial neural network with nine weights. However, the procedure can find solutions for any number of dimensions. The results achieved demonstrate the viability of the proposal, contributing to machine learning and quantum computing researches.

\keywords{Artificial Neural Networks Training \and Quantum Computing \and Quantum Walk \and Search Algorithm.}
\end{abstract}
\section{Introduction}
\label{sec:introduction}


The need for increased computing power and the miniaturization of components at scales where quantum effects cannot be ignored \citep{powell2008quantum} support the idea that quantum physics can redefine a new frontier for computing problems by taking an essential role in the computational race \citep{singh2016evolution}. Quantum effects can provide computational gains and algorithms that are mostly more efficient than their classical counterparts for many problems. Quantum computing seeks through the use of aspects of quantum mechanics to expand computational horizons \citep{lloyd2013quantum,yanofsky2008quantum}. An example of the computational power of quantum computing is performing a search for elements in a disordered database in just $\mathbf{O}(\sqrt{N})$ \citep{grover1996fast}, where $N$ is the size of the database. The gain is quadratically faster than its classical analog.

In this perspective, efforts are employed in the search for more efficient algorithms. In particular, there is a branch of research called quantum machine learning (QML) \citep{wittek2014quantum}, where machine learning methods and artificial intelligence are integrated into the world of quantum computing in the hope to find more efficient procedures than classical algorithms. According to \citet{dunjko2016quantum}, the quantum processing of information from quantum machine learning is interactively configured in two parts called the agent-environment. These parts are grouped into four categories:  $CC$, $CQ$, $QC$, and $QQ$, where  $C$  means \textit{Classical}, $Q$ means \textit{Quantum}. The first letter is referent to the agent and the second to the environment.  $CC$ encompasses classical machine learning. $CQ$ analyzes how classical learning techniques can help in quantum tasks. $QC$ represents the quantum variants of classical learning algorithms facing a classical environment, where this work is classified. $QQ$ is the quantum systems world.


Papers grouped within the $QC$ category are developed to improve classical procedures with the quantum information processing (QIP) paradigm. \citet{aimeur2013quantum} showed that it is possible to accelerate unsupervised learning algorithms by quantizing some of their subroutines. \citet{zheng2018quantum} designed an algorithm to train a perceptron using Grover's Algorithm \citep{grover1996fast}. \citet{schuld2015simulating} developed a procedure that uses a quantum phase estimation algorithm in the classical neural network training process. \citet{date2020classical} presents a Classical-Quantum Hybrid Approach for training unsupervised probabilistic machine learning models.


Machine learning (ML) studies techniques to give machines the ability to learn from past experiences. Its core tasks include classification or regression in supervised learning and density estimation or clustering in unsupervised learning, for example. Usually, in the ML context, the training of a machine is performed using a learning algorithm that uses as input a training data set \citep{aimeur2006machine} to extract information, adjust its parameters, and solve a given problem.

One of the ML techniques widely employed for many problems is the Artificial Neural Network (ANN), wherein the classical (non-quantum) world is a Classical ANN.  Classical ANNs are one of the best-known classifiers and predictors. Classical ANNs have proven to be very competitive in solving real-world problems compared to other conventional data analysis methodologies \citep{prieto2016neural}. Its optimization is observed from various perspectives, but in general, classical ANN training is performed mainly using a gradient descent algorithm. However, optimization methods based on the descending gradient algorithm have limitations. Aspects such as weight initialization, network architecture, activation functions, meta parameters, and learning environment can influence the optimization process \citep{haykin,ojha2017metaheuristic}. In this way, the training algorithm may not obtain a set of weights that train the neural network, staying stagnated at local minimums.


Therefore, a neural network's training algorithm based on the descending gradient method can be seen as a search problem and seeks to minimize an error function.  This ANN training is a search problem for an appropriate weight configuration that allows learning the network \citep{biamonte2017quantum}. On the other hand, in the quantum computing branch, some search algorithms are more efficient than their classical analogue. For example, Grover's quantum search algorithm \citep{grover1996fast} and quantum walks \citep{wong2015grover,nahimovs2015exceptional,wong2018faster,lovett2019quantum,nahimovs2019lackadaisical}. In this perspective,  quantum search algorithms have already been used for the training process of classical ANN \citep{schuld2015simulating,zheng2018quantum}.

Based on the incipient work developed by \citet{souza2019quantum} with extreme learning machines, this article generalizes and extends the quantum walk proposal to train a classical ANN. Here, a full graph lackadaisical quantum walk algorithm \citep{wong2015grover,wong2018faster,nahimovs2019lackadaisical} is applied as a search method to find all the synaptic weights that optimize the learning procedure of a classical ANN.


This paper is organized as follows. Section \ref{sec:quantum-walk} introduces some concepts of the one-dimensional and the generalization for $n$-dimensional quantum walks, and it also presents the quantum walk on a complete graph. Section \ref{sec:proposal-procedure}, it shows the computational procedure used in this paper. Section \ref{sec:experiment-setup}, it shows the experiments performed. Section \ref{sec:experimental-results-and-discussions} discusses the results obtained. Finally, Section \ref{sec:conclusion} is the conclusion of the work.
\section{Quantum Walk}
\label{sec:quantum-walk}

The simplest model of the classical random walk can be described by a particle's classical movement in a straight line \citep{portugal2013quantum}. Let the $s$ the particle probability of going to the right. Let $(1 - s)$ the probability of going to left. Therefore, the direction of the particle is conditioned by tossing a coin. This process is probabilistic, so it is impossible to know with certainty where the particle will be at any given time. However, it is possible to calculate the probability $p$ that particle is at a point $n$ at time t, as shown in Equation \ref{eq:probability-of-the-classic-walk} for the case $s = 1/2$.

\begin{equation}\label{eq:probability-of-the-classic-walk}
    p(t,n) \simeq \frac{2}{\sqrt{2 \pi t}} e^{-\frac{n^2}{2t}}
\end{equation}

The quantum walks generalize the concept of a classical random walk, \textit{i.e.}, the quantum mechanical counterpart of classical random walks \citep{venegas2012quantum}. It is assuming a walker represented by a normalized vector in Hilbert space. The quantum walk evolves in the Hilbert space $\mathcal{H}_{M} \otimes \mathcal{H}_{P}$, where $\mathcal{H}_{M}$ is the coin space that controls the walker's movement, and $\mathcal{H}_{P}$ defines the position of the walker \citep{portugal2013quantum}.

Suppose the quantum walk takes place in a one-dimensional space. In this case, the coin space needs two degrees of freedom. Therefore, the coin space $\mathcal{H}_{M}$ is generated by the computational base $\{\ket{0}, \ket{1}\}$. A qubit can represent this quantum information. The possible states that one qubit can assume are represented by the state vectors $\ket{0}$ and $\ket{1}$ described in Equations \ref{eq:representacao-quibts-ket-zero} and \ref{eq:representacao-quibts-ket-one}, respectively.

\begin{equation}
\label{eq:representacao-quibts-ket-zero}
\ket{0} = \begin{bmatrix} 1 \\ 0 \end{bmatrix}\end{equation}

\noindent and

\begin{equation}
\label{eq:representacao-quibts-ket-one}
\ket{1} = \begin{bmatrix} 0 \\ 1 \end{bmatrix}\end{equation}

Walker space $\mathcal{H}_{P}$ is generated by base $\{\ket{n} : n \in \mathbb{Z}\}$ which represents all integers of one-dimensional space. Consider an operator $S$ that, when applied to the system formed here by a coin and a walker, will shift the position of the individual to $\ket{n + 1}$ or $\ket{n - 1}$ depending on the state of the coin according to Equation \ref{eq:operator-s}.

\begin{equation}\label{eq:operator-s}
\begin{matrix}
S\ket{0}\ket{n} = \ket{0}\ket{n + 1}\\\\
S\ket{1}\ket{n} = \ket{1}\ket{n - 1}
\end{matrix}
\end{equation}

A unitary transformation describes the evolution of a closed quantum system. This evolution depends on the application of an operator $U$ shown in Equation \ref{eq:evolution-operator} to the system over time \citep{nielsen2002quantum},

\begin{equation}
    \label{eq:evolution-operator} U = S(H \otimes I)
\end{equation}
where $H$ is the Hadamard operator and $I$ is the identity.

For an operator $U$ of a quantum system to be unitary, it must satisfy the condition described in Equation \ref{eq:condicao-unitariedade}, where $U^{\dagger}$ is the adjunct of $U$. This condition is necessary for the norm of the vectors to be maintained, for this we must calculate $\parallel u \parallel \textrm{ = } \sqrt{\braket{u\mid u}}$, where $\braket{u\mid u}$ is the inner product.

\begin{equation}\label{eq:condicao-unitariedade}
    UU^{\dagger} = U^{\dagger}U = I
\end{equation}

\noindent The evolution of the quantum walk system is performed by the  Equation \ref{eq:system-state-at-time-t}.

\begin{equation}\label{eq:system-state-at-time-t}
\ket{\Psi(t)} = U^{t}\ket{\Psi(0)}
\end{equation}
where $\ket{\Psi(0)}$ is the initial state of the quantum system.

\begin{figure}
    \centering
    \includegraphics[height=6cm]{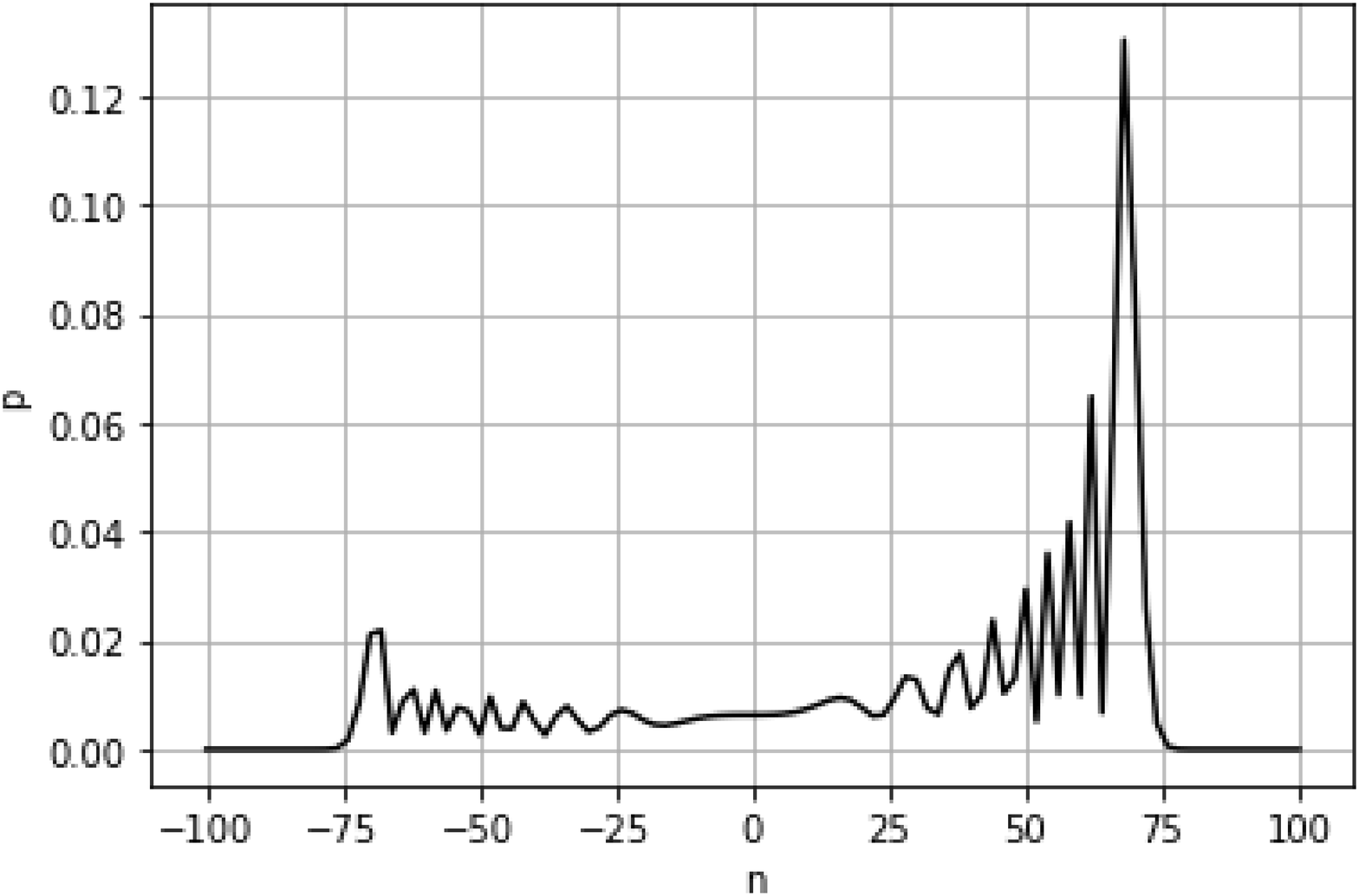}
    \caption{Probability distribution of the one-dimensional quantum walk after 100 steps. The initial state is described by the Equation \ref{eq:right-initial-condition}. The points with zero value were ignored.}
    \label{fig:probability-distribution-of-asymmetric-one-dimensional-walk}
\end{figure}

\begin{figure}
    \centering
    \includegraphics[height=6cm]{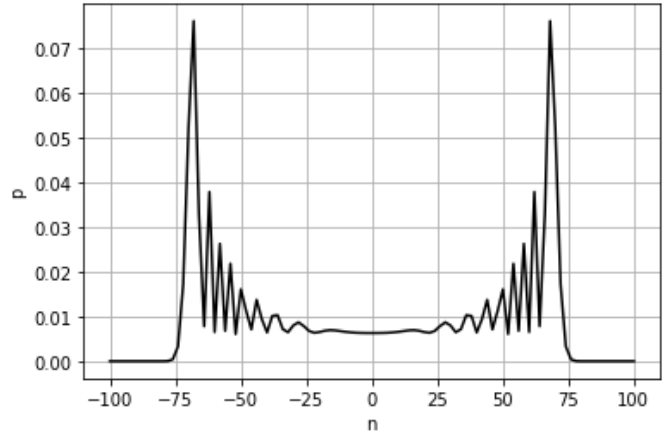}
    \caption{Probability distribution of the one-dimensional quantum walk after 100 steps. The initial state is described by the Equation \ref{eq:symmetrical-initial-condition}. The points with zero value were ignored.}
    \label{fig:probability-distribution-of-symmetrical-one-dimensional-walk}
\end{figure}

Consider the initial state of the quantum system given by the Equations \ref{eq:right-initial-condition} and \ref{eq:symmetrical-initial-condition}. Asymmetrical and symmetrical state, respectively. We can obtain the asymmetrical and symmetrical probability distributions after one hundred applications of $U$ operator, shown in Figures \ref{fig:probability-distribution-of-asymmetric-one-dimensional-walk} and \ref{fig:probability-distribution-of-symmetrical-one-dimensional-walk}. Unlike the classical case, where the distribution is an origin-centered Gaussian, for both case asymmetrical and symmetrical, the quantum walk has a large spread with an interval of $-t/\sqrt{2}$ to $t/\sqrt{2}$.

\begin{equation}
\label{eq:right-initial-condition}
\ket{\Psi(0)} = \ket{0}\ket{n = 0}
\end{equation}

\begin{equation}
\label{eq:symmetrical-initial-condition}
\ket{\Psi(0)} = \frac{\ket{0}-i\ket{1}}{\sqrt{2}}\ket{n=0}
\end{equation}

For example, if the quantum walk starts in the initial state described in Equation \ref{eq:right-initial-condition}. Apply the Hadamard operator $H$ as the coin of the quantum walk system and then applies the shift operator $S$. We have the initial evolution state presented in Equation \ref{eq:asymmetrical-walk-execution-first-step} at the end of the first step. The Hadamard operator application in the computational base vectors generates a superposition state, \textit{i.e}, one of the most well-known quantum effects that qubits are at the same time in distinct states.

\begin{align}\label{eq:asymmetrical-walk-execution-first-step}
\begin{split}
\ket{\Psi(1)} &= \frac{1}{\sqrt{2}}(\ket{0}\ket{1} + \ket{1}\ket{-1})
\end{split}
\end{align}

With the successive application of the evolution operator $U$ described in Equation \ref{eq:evolution-operator}, at the end of the third stage, we can observe in Equation \ref{eq:asymmetrical-walk-execution-second-third-step} that the state $\ket{\Psi(3)}$ is asymmetrical about the origin. This asymmetry will keep for all system evolution, as can be observed in Figure \ref{fig:probability-distribution-of-asymmetric-one-dimensional-walk}.

\begin{align}\label{eq:asymmetrical-walk-execution-second-third-step}
\begin{split}
\ket{\Psi(2)} &= \frac{1}{2}(\ket{0}\ket{2} + (\ket{1} + \ket{0})\ket{0} - \ket{1}\ket{-2})\\
\ket{\Psi(3)} &= \frac{1}{2\sqrt{2}}(\ket{0}\ket{3} + (2\ket{0} + \ket{1})\ket{1} \\&- \ket{0}\ket{-1} + \ket{1}\ket{-3})
\end{split}
\end{align}

\subsection{One-dimensional quantum walk}
\label{sec:one-dimensional-quantum-walk}

This section will describe the one-dimensional quantum walk model analytically and recursively. The generic state for a one-dimensional quantum walk model is described in Equation \ref{eq:generic-state-of-walk} \citep{portugal2013quantum}.

\begin{equation}\label{eq:generic-state-of-walk}
    \ket{\Psi(t)} = \sum_{n=-\infty}^{\infty} (\alpha_{n}(t)\ket{0} + \beta_{n}(t)\ket{1})\ket{n},
\end{equation}
where the coefficients $\alpha_n(t)$ and $\beta_n(t)$ satisfy the condition described in Equation \ref{eq:one-dimensional-initial-condition}.

\begin{equation}\label{eq:one-dimensional-initial-condition}
    \sum_{n=-\infty}^{\infty} \left | \alpha_{n}(t) \right |^{2} + \left | \beta_{n}(t) \right |^{2} = 1
\end{equation}
Applying the operator $H \otimes I$ to state $\ket{\Psi(t)}$, we find recursive formulas involving the coefficients $\alpha$ and $\beta$ in Equations \ref{eq:recursive-formula-alpha} and \ref{eq:recursive-formula-beta}.

\begin{equation}
\label{eq:recursive-formula-alpha}
    \alpha_{n}(t+1)=\frac{\alpha_{n-1}(t)+\beta_{n-1}(t)}{\sqrt{2}}
\end{equation}

\begin{equation}
\label{eq:recursive-formula-beta}
   \beta_{n}(t+1)=\frac{\alpha_{n+1}(t)-\beta_{n+1}(t)}{\sqrt{2}}
\end{equation}

As already shown in Figures \ref{fig:probability-distribution-of-asymmetric-one-dimensional-walk} and \ref{fig:probability-distribution-of-symmetrical-one-dimensional-walk}, the quantum walk probability distributions are dependent on the initial state. Therefore, the initial state of the system can generate a walk with symmetrical or asymmetrical probability distribution about the origin. The probability distribution can be calculated using the Equation \ref{eq:probability-distribution-calculation}.

\begin{equation}\label{eq:probability-distribution-calculation}
    p(t,n) = \left | \alpha_{n}(t) \right |^{2} + \left | \beta_{n}(t) \right |^{2}
\end{equation}

\subsection{n-dimensional quantum walk}
\label{sec:n-dimensional-quantum-walk}

It is possible to generalize the concept presented in Section \ref{sec:one-dimensional-quantum-walk} to any number of dimensions. Consider the quantum walk in an infinite n-dimensional grid with the associated Hilbert space $\mathcal{H}_{M} \otimes \mathcal{H}_{P}$, whose $\mathcal{H}_{P}$'s computational base is $\{\ket{x,y,\dots,n} : x,y,\dots,n \in \mathbb{Z}\}$,  and the Coin space $\mathcal{H}_{M}$'s computational basis is $\{\ket{i_{x},i_{y},\dots,i_{n}} :  i_{x},i_{y},\dots,i_{n} \in \{0,1\}^{n}\}$. The generic state for this quantum walk model at time $t$ is presented in Equation \ref{eq:equacao-estado-inicial-n-dimensional}.

\begin{equation}
\label{eq:equacao-estado-inicial-n-dimensional}
    \ket{\Psi(t)} = \sum_{i_{x},i_{y},\dots,i_{n} = 0}^{1} \sum_{x,y,\dots,n = -\infty}^{\infty} \psi_{i_{x},i_{y},\dots,i_{n};x,y,\dots,n}(t) \ket{i_{x},i_{y},\dots,i_{n}}\ket{x,y,\dots,n}
\end{equation}

\noindent where $\psi_{i_{x},i_{y},\dots,i_{n};x,y,\dots,n}(t)$ are complex functions that satisfy the condition shown in Equation \ref{eq:equacao-condicao-de-normalizacao2d} for all time $t$.

\begin{equation}
    \label{eq:equacao-condicao-de-normalizacao2d}
    \sum_{i_{x},i_{y},\dots,i_{n} = 0}^{1} \sum_{x,y,\dots,n = -\infty}^{\infty} \left | \psi_{i_{x},i_{y},\dots,i_{n};x,y,\dots,n}(t) \right |^{2} = 1
\end{equation}
Moreover, it is possible to calculate the distribution of probabilities using Equation \ref{eq:equacao-distribuicao-de-probabilidade}.

\begin{equation}
    \label{eq:equacao-distribuicao-de-probabilidade}
    p_{x,y,\dots,n}(t) = \sum_{i_{x},i_{y},\dots,i_{n} = 0}^{1} \left | \psi_{i_{x},i_{y},\dots,i_{n};x,y,\dots,n}(t) \right |^{2}
\end{equation}

Applying the standard evolution operator $U = S(C \otimes I)$ ($C$ is the coin operator) to the generic state described in Equation \ref{eq:equacao-estado-inicial-n-dimensional} and making the expansions, we obtain Equation \ref{eq:equacao-de-evolucao-n-dimensional} which is the walker evolution equation \citep{portugal2013quantum}.

\begin{equation}
\label{eq:equacao-de-evolucao-n-dimensional}
\psi_{i_{x},i_{y},\dots,i_{n};x,y,\dots,n}(t+1) = \sum_{j_{x},j_{y},\dots,j_{n} = 0}^{1} C_{i_{x},i_{y},\dots,i_{n};j_{x},j_{y},\cdots,j_{n}}\\ \psi_{j_{x},j_{y},\dots,j_{n};x+(-1)^{i_{x}},y+(-1)^{i_{y}},\dots,n+(-1)^{i_{n}}}(t)
\end{equation}

\subsection{Lackadaisical Quantum Walk on Complete Graph}
\label{sec:lackadaisical-quantum-walk-in-complete-graph}

\begin{figure}
\centering
\includegraphics[height=7cm]{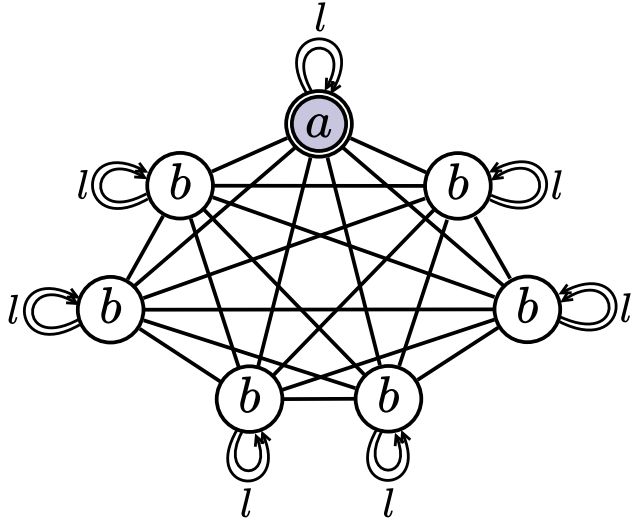}
\caption{Complete graph with $N=7$ vertices. The single vertex marked as $a$ is indicated by the double circle shaded. Adapted from Wong's work \citep{wong2015grover}. Grover search with lackadaisical quantum walks.}
\label{fig:complete-graph-with-selfloops}
\end{figure}

The quantum walks presented in the previous sections are the basis for other techniques that can be used in other search spaces. One such variation is the quantum walk in a complete graph developed by \citet{wong2015grover}, represented in Figure \ref{fig:complete-graph-with-selfloops}.

There are two types of vertex, $a$ and $b$, marked as a solution and non-solution, respectively. Each one vertex has $l$ self-loops. This approach considers the walker's movement into the complete graph to create the states of the new computational basis.

If the walker is on an $a$ vertex, there are two movement options. It can move to a vertex that is a solution ($a \rightarrow a $) or to a vertex that is not a solution ($ a \rightarrow b $). Defining the quantum states $ \ket{a} \otimes \ket{a \rightarrow a'} $ and $ \ket{a} \otimes \ket{a \rightarrow b} $. Similarly, if the walker is on any vertex $b$. It can move to a vertex that is a solution ($b \rightarrow a$) or to a vertex that is not a solution ($b \rightarrow b'$). Defining the quantum states $ \ket{b} \otimes \ket{b \rightarrow a}$ and $ \ket{b} \otimes \ket{b \rightarrow b'}$.

Note that the states $\ket{a}$ or $\ket{b}$ represent the graph vertices and $\ket{a \rightarrow b}$, for example, represents the edges where $\ket{a}$ is the walker's current state and $\ket{b}$ is the state to which the walker will move to it. Equation \ref{eq:definicao-base-quatro-estados} shows the states $\ket{AA}, \ket{AB},\ket{BA}, \textrm{and} \ket{BB}$ of the new quantum states for the situation where there is only one solution (only one vertex labeled as $a$). $N$ is the total number of vertices. Since there is only one solution, $k = 1$, and the number of self-loops is greater than zero, $l > 0$.

\begin{align}
\begin{split}
\label{eq:definicao-base-quatro-estados}
    \ket{AA} &=  \frac{1}{\sqrt{l}} \ket{a} \otimes \ket{a \rightarrow a'} \\
    \ket{AB} &= \frac{1}{\sqrt{N-1}} \sum_{b} \ket{a} \otimes \ket{a \rightarrow b} \\
    \ket{BA} &=  \frac{1}{\sqrt{N-1}} \sum_{b} \ket{b} \otimes \ket{b\rightarrow a}\\
    \ket{BB} &= \frac{1}{\sqrt{(N-1)(N+l-2)}} \sum_{b} \sum_{b'} \ket{b} \otimes \ket{b \rightarrow b'}
\end{split}
\end{align}

Equation \ref{eq:estado-inicial-grafo-completo} defines the initial state  $\ket{\Psi_{0}}$. This state is described as the uniform superposition $N^{-1} \sum_{x,y} \ket{x}\ket{x \rightarrow y}$ expressed in terms of the states $AA, AB, BA$ and $BB$.

\begin{align}
\small
\begin{split}
    \label{eq:estado-inicial-grafo-completo}
    \ket{\Psi_{0}} &= \frac{1}{\sqrt{N(N+l-1)}} ( \sqrt{l} \ket{AA} + \sqrt{N-1} \ket{AB}\\
    & + \sqrt{N-1} \ket{BA} + \sqrt{(N-1)(N+l-2)} \ket{BB})
\end{split}
\end{align}

The lackadaisical quantum walk is accomplished by successive applications of a unitary operator $U$, defined in Equation \ref{eq:operador-evolucao-grafo-completo} which inverts the sign of solution states $\ket{a}\ket{a \rightarrow x}$ using an oracle and swaps vertices on each edge, $\ket{x}\ket{x \rightarrow y} \rightarrow \ket{y}\ket{y \rightarrow x}$ as described in \citet{wong2015grover}.

\begin{equation}
\label{eq:operador-evolucao-grafo-completo}
 U = \begin{pmatrix}
\cos{\theta} & -\sin{\theta}  & 0 & 0 \\ 
0 & 0 & -\cos{\phi} & \sin{\phi} \\ 
-\sin{\theta} & -\cos{\theta} & 0 & 0 \\ 
0 & 0 & \sin{\phi} & \cos{\phi} 
\end{pmatrix}
\end{equation}
where $\theta$ is defined such that,

\begin{align*}
    \cos{\theta} = \frac{N-l-1}{N+l-1} 
\end{align*}
and

\begin{align*}
    \sin{\theta} = \frac{2\sqrt{l(N-1)}}{N+l-1} 
\end{align*}
and $\phi$ is defined such that,

\begin{align*}
    \cos{\phi} = \frac{N-l-3}{N+l-1}
\end{align*}
and

\begin{align*}
    \sin{\phi} = \frac{2\sqrt{N+l-2}}{N+l-1} 
\end{align*}

Thus, the evolution of the system occurs in a four-dimensional subspace, and each state of the new representation is formed by overlapping vertices and edges \citep{wong2015grover}.

\subsubsection{Quantum walk with self-loops for k solutions}
\label{sec:quantum-walk-with-self-loops-for-k-solutions}

The previous section introduced the quantum walk in a complete graph and describes the approach to the case of a single solution ($k = 1$) and the number of self-loops $l > 0$. Now, consider a number of solutions  $k > 1$ and only one self-loop ($l = 1$) per vertex.

The solution set has the number $k$ of vertices, and the non-solution set has $N-k$ vertices, where $N$ is the total number of vertices. We will use the previous idea that considered the walker's movement to define the states of the new multiple solutions computational basis.

If the walker is in a vertex $a$ moving to another vertex $a'$, there will exist $k$ edges of the type $\ket{a\rightarrow a'}$. If the walker is in a vertex $a$ moving to a vertex $b$, there will exist $N - k$ edges of the type $\ket{a\rightarrow b}$. Similarly, if the walker is in a vertex $b$ moving to a vertex $a$, there will exist $k$ edges of the type $\ket{b\rightarrow a}$. And if walker is in a vertex $b$ moving to another vertex $b'$, there will exist $N-k$ edges of the type  $\ket{b\rightarrow b'}$.

The number of vertices marked as solution must be of the order $o(N)$ because if  $k = O(N)$, then $k = cN$ (in the limit of $N\rightarrow \infty$ and $c$ is a finite constant). In this last case, the search for one solution could be performed classically in an efficient way,  \textit{i.e.}, in a constant number of guesses \citep{wong2015grover}. Thus, the new quantum states are redefined in Equation \ref{eq:redefinicao-base-quatro-estados}.

\begin{align}
\begin{split}
\label{eq:redefinicao-base-quatro-estados}
\ket{AA} &= \frac{1}{k} \sum_{a}^{} \sum_{a'}^{}
\ket{a} \otimes \ket{a \rightarrow a'} \\
\ket{AB} &= \frac{1}{\sqrt{k(N-k)}}  \sum_{a}^{} \sum_{b}^{} \ket{a} \otimes \ket{a \rightarrow b} \\
\ket{BA} &= \frac{1}{\sqrt{k(N-k)}}  \sum_{b}^{} \sum_{a}^{} \ket{b} \otimes \ket{b \rightarrow a} \\
\ket{BB} &= \frac{1}{N-k} \sum_{b}^{} \sum_{b'}^{} \ket{b} \otimes \ket{b \rightarrow b'}
\end{split}
\end{align}

The initial state of the system is rewritten, as presented in Equation \ref{eq:estado-inicial-grafo-completo-redefinido}.

\begin{equation}
    \label{eq:estado-inicial-grafo-completo-redefinido}
    \ket{\Psi_{0}} = \frac{1}{N} (k \ket{AA} + \sqrt{k(N-k)} \ket{AB} + \sqrt{k(N-k)} \ket{BA} + (N-k) \ket{BB})
\end{equation}

A modification to the evolution operator $U$ (Equation \ref{eq:operador-evolucao-grafo-completo}) is made so that the number of solutions is included in the definitions of $\theta$ and $\phi$ \citep{wong2015grover}. Therefore, $\theta$ is redefined according to Equations \ref{eq:redefinicao-theta-cos} and \ref{eq:redefinicao-theta-sin},

\begin{align}
\label{eq:redefinicao-theta-cos}
    \cos{\theta} = \frac{N-2k-l+1}{N+l-1}
\end{align}

\begin{align}
\label{eq:redefinicao-theta-sin}
    \sin{\theta} = \frac{2\sqrt{(N-k)(k+l-1)}}{N+l-1} 
\end{align}

\noindent and $\phi$ is redefined according to Equations \ref{eq:redefinicao-phi-cos} and \ref{eq:redefinicao-phi-sin},

\begin{align}
\label{eq:redefinicao-phi-cos}
    \cos{\phi} = \frac{N-2k+l-1}{N+l-1} 
\end{align}

\begin{align}
\label{eq:redefinicao-phi-sin}
    \sin{\phi} = \frac{2\sqrt{k(N-k+l-1)}}{N+l-1} 
\end{align}

The maximum success probability value is reached after the number of steps $t$ defined in Equation \ref{eq:qtd-passos-probabilidade-maxima} \citep{wong2015grover}. Success is defined as the measurement of some state $\ket{AA}$ or the state $\ket{AB}$. Both states represent the set of vertices marked as a solution.

\begin{align}
    \begin{split}
    \label{eq:qtd-passos-probabilidade-maxima}
    t = \frac{\pi}{\sqrt{2(2k+l-1)}}\sqrt{N}
    \end{split}
\end{align}

\subsubsection{A Toy Example}
\label{sec:toy-problem}

Consider the particular case of the lackadaisical quantum walk in a complete graph with $N = 8$ vertices, $k = 2$ solutions, and $l = 1$ self-loops at each vertex. For illustrative purposes, these vertices have also been marked with sub-indices, so $a_1$ and $a_2$ are the solutions whereas vertices $b_1,\dots,b_6$ are not solutions. In practice, however, sub-indices or whatever kind of information that can distinguish the solutions from one another, or the non-solutions from one another, are not available. The vertices are marked only with $a$ or $b$, exclusively.

The quantum state $\ket{AB}$, for example, is formed by all vertices that are solution combined with their respective edges for non-solution vertices. Therefore, the state $\ket{AB}$ for the given example is defined according to Equation \ref{eq:estado-ab-toy-problem}.

\begin{align}
\begin{split}
\footnotesize
\label{eq:estado-ab-toy-problem}
\ket{AB} = \frac{1}{\sqrt{k(N-k)}} \sum_{a}^{} \sum_{b}^{} \ket{a} \otimes \ket{a \rightarrow b} 
= \frac{1}{\sqrt{12}} \bigg[ \Big( \ket{a_1}\ket{a_1\rightarrow b_1} + \ket{a_1}\ket{a_1\rightarrow b_2}
+ \ket{a_1}\ket{a_1\rightarrow b_3}\\
+ \ket{a_1}\ket{a_1\rightarrow b_4}
+ \ket{a_1}\ket{a_1\rightarrow b_5}
+ \ket{a_1}\ket{a_1\rightarrow b_6} \Big)
+ \Big( \ket{a_2}\ket{a_2\rightarrow b_1}
+ \ket{a_2}\ket{a_2\rightarrow b_2}\\
+ \ket{a_2}\ket{a_2\rightarrow b_3}
+ \ket{a_2}\ket{a_2\rightarrow b_4}
+ \ket{a_2}\ket{a_2\rightarrow b_5}
+ \ket{a_2}\ket{a_2\rightarrow b_6} \Big) \bigg]
\end{split}
\end{align}

The analysis for the quantum states $\ket{AA}$, $\ket{BA}$ and $\ket{BB}$ is analogous. With the new quantum states prepared, the initial state of the system $\ket{\Psi_{0}}$ can be defined according to Equation \ref{eq:estado-inicial-toy-problem}. This initial state is also normalized. 

\begin{equation}
\small
    \label{eq:estado-inicial-toy-problem}
    \ket{\Psi_{0}} = \frac{1}{8} \left( 2 \ket{AA} + \sqrt{12} \ket{AB} + \sqrt{12} \ket{BA} + 6 \ket{BB} \right)
\end{equation}

Making only three ($t = 3$) successive applications of the evolution operator $U$, represented by the matrix of Equation \ref{eq:operador-evolucao-toy-problem}, the probability of success (the measurement of a solution state) tends to $1$. A measurement made in $\ket{\Psi_{3}}$ makes the system to collapse to the state $\ket{AA}$. All energy of the system is concentrated at this state $\ket{AA}$.

\begin{equation}
\label{eq:operador-evolucao-toy-problem}
U = \frac{1}{2} \begin{pmatrix}
1 & -\sqrt{3} & 0 & 0 \\ 
0 & 0 & -1 & \sqrt{3} \\ 
-\sqrt{3} & -1 & 0 & 0 \\ 
0 & 0 & \sqrt{3} & 1
\end{pmatrix}
\end{equation}
\section{Proposal Procedure}
\label{sec:proposal-procedure}

Quantum walks are algorithms that can be applied to search problems \citep{shenvi2003quantum,lovett2019quantum,wong2018faster}, where the proof of its correctness can be found in \citep{feng2007proof}. Thus, it is possible to idealize this algorithm's application to find the set of synaptic weights that train a classical artificial neural network. Based on the concept of quantum information processing called agent-environment presented in Section \ref{sec:introduction}, the objective is to replace the classical algorithm with a quantum search algorithm in a neural network training process. 

In a classical environment, we replace the backpropagation algorithm for a quantum walk to search synaptic weights. The evolution of the quantum walk occurs by applying a unitary operator $U$ over an initial state. Subsequently, a measurement of the walker state is performed. If the measurement is performed at each step, then the quantum walk falls in the classical case. In this situation, the correlations between the different positions of the walker are lost.

For the correlations between the walker's positions to be maintained, the measurement should not occur at every single step.  Thus,  the measurement process only will occur after a predetermined time $t$. Once the correlations between positions hold, constructive and destructive interference occurs \citep{portugal2013quantum}. The interference caused by these relationships between positions generates the probability amplitudes for each position. Depending on the initial state of the system, the probability distribution may be asymmetrical or symmetrical according to equations \ref{eq:right-initial-condition} and \ref{eq:symmetrical-initial-condition}, respectively, as viewed in Section~\ref{sec:quantum-walk}. As observed in Figures \ref{fig:probability-distribution-of-asymmetric-one-dimensional-walk} and \ref{fig:probability-distribution-of-symmetrical-one-dimensional-walk}, according to the probability distribution, the chances of finding the walker on the extremities are higher than finding it in another position. In this way, it is possible to obtain these extreme positions of the walker with a high probability of measurement.

Therefore, it is possible to determine with a high probability the walker's extreme position after a given number of steps. If this position does not contain a solution, the quantum walk search will not succeed. It is necessary to know the position where there is a solution and guarantee an amplitude amplification for this position. It is also necessary to guarantee that the solution state at the time of measurement has a high probability of measurement. The complete-graph quantum walk proves to be adequate for these proposals. The complete-graph quantum walk employed here considers only the information about the \textit{solution} or \textit{no solution} label ($a$ or $b$) contained in all vertices to execute the quantum walk evolution. However, to define these labels, the information about the weights used to train the classical ANN is necessary. Thus, besides the label for \textit{solution} ($\ket{a}$) and \textit{no solution} ($\ket{b}$), each vertex will also have the information about the associated weights used to train the classical ANN. 

\begin{figure}
\centering
\includegraphics[width=0.9\textwidth]{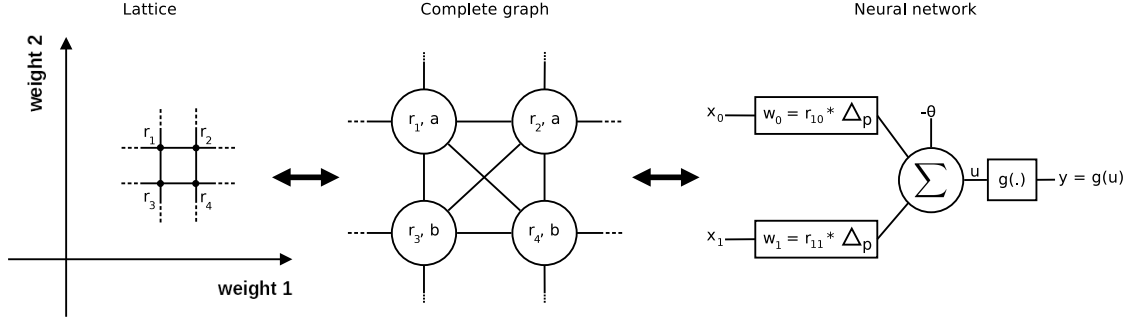}
\caption{On the left side, a discrete schematic representation of the set of weight values of an artificial neural network. This representation generates a grid, represented here by a lattice. Each intersection point $i$ of the lattice has the information about the set of weight values $r_{i}$. With an oracle operator, the state solution label ($a$ for a solution and $b$ for no solution) is determined. On the central figure, the complete graph representation of the grid. On the right side, a neural network with one neuron demonstrates the relationship between the labels on the grid, the complete graph, and the definition of synaptic weights w$_0$ and w$_1$, where $g(\cdot)$ is the activation function, and $u = x_0 \cdot $w$_0 + x_1\cdot $w$_1 - \theta$.} \label{fig:Lattice}
\end{figure}

Consider a $w$-dimensional lattice.  Let a classical ANN with the number of weights equal to the lattice's dimension, $w$ weights. In the discrete representation, each intersection of lattice lines can represent an ANN $w$ weights configuration. Figure ~\ref{fig:Lattice} represents this idea for the $2$-dimensional case. In this way, each point $i$ in this $w$-dimensional lattice will be a vertex in a complete graph, where it was labeled as $a$ if its value configuration of weights $r_i$ is a solution for the ANN. Otherwise, it was labeled as $b$. The labels $a$ and $b$ are created by applying the oracle to the grid. Change the grid representation to the complete graph creates a new representation, where all $i$ grid point is a vertex with the information $r_i$ (the weight set) and the label $a$ or $b$. Wong's quantum walk \citep{wong2015grover} is applied in the complete graph, where now each vertex has the label $a$ or $b$, and the associated weight set ($r_i$) used to train the classical ANN. Then, the search procedure is done, where a vertex labeled by $\ket{a}$ is searched.

Thus, to recover the synaptic weights that trained the ANN after the quantum walk evolution, it is necessary to obtain the specific weight values configuration $r_i$ of the state measured. For this reason, a modification is proposed to the original procedure, including the weight vector $\ket{\vec{r_i}}$ in the base states definition.  All quantum walk search procedure is the same, where the state with label $a$ is sought. However, with this new associated information about the weight vector $\ket{\vec{r_i}}$, it is possible to determine the walker's position on the grid (or lattice) in the final measurement procedure in the final state found by the quantum walk algorithm. In this way, the information about the set of weight values used to train the classical ANN can be recovered. The new definition of states is presented in Equation \ref{eq:redefinicao-base-quatro-estados-rotulo}. With this modification, it is proposed the search procedure presented in Algorithm~\ref{alg:algoritmo-em-grafo-completo}.

\begin{align}
\begin{split}
\label{eq:redefinicao-base-quatro-estados-rotulo}
\ket{AA} &= \frac{1}{k} \sum_{a}^{} \sum_{a'}^{}
\ket{\vec{r}_{a}} \ket{a} \otimes \ket{a \rightarrow a'} \\
\ket{AB} &= \frac{1}{\sqrt{k(N-k)}} \sum_{a}^{} \sum_{b}^{} \ket{\vec{r}_{a}} \ket{a} \otimes \ket{a \rightarrow b} \\
\ket{BA} &= \frac{1}{\sqrt{k(N-k)}} \sum_{b}^{} \sum_{a}^{} \ket{\vec{r}_{b}} \ket{b} \otimes \ket{b \rightarrow a} \\
\ket{BB} &= \frac{1}{N-k} \sum_{b}^{} \sum_{b'}^{} \ket{\vec{r}_{b}} \ket{b} \otimes \ket{b \rightarrow b'}
\end{split}
\end{align}

Initially, it is necessary to define some initial parameters of the Algorithm \ref{alg:algoritmo-em-grafo-completo}. A $w$-dimensional grid (or a lattice) will define the possible weight configurations for the walker, where each dimension represents an ANN weight set. Therefore, the user defines the number of grid points, $N$, and the distance between adjacent points, $\Delta p$. In this way, $N$ will be the number of possibles positions state of the system for the walker, represented by the complete graph, and $\Delta p$ will define the granularity of the weights representation. If $\Delta p$ is small, the weight representation will have high resolution, but a significant $N$ value is necessary to search for a solution in practice.  If $\Delta p$ is large, the weight search will have low resolution, but a small $N$ value is sufficient to cover a given search space. The algorithm employs an oracle, which defines the vertices that are solution and non-solution. The oracle employed here is described in Section \ref{sec:representation-oracle}. In this way, it is considered that the oracle used in this work is independent of the algorithm proposed. An $w$-dimensional sparse matrix with $N$ elements represents classically the oracle, where the element marked with number one denotes a solution and with zero a non-solution. The user also defines the number of self-loops per vertex, $l$. However, the number of self-loops per vertex here always was $l = 1$ for all experiments.

After defining the initial parameters, the proposed algorithm performs a quantum count to estimate the number of solutions $k$ in the search space, as indicated in line $3$ of the Algorithm \ref{alg:algoritmo-em-grafo-completo}. It is necessary to know the quantity $k$ of solutions to determine the initial state, the shift operator $U$, and the number $t$ of iterations. It is possible to define the number of solutions by combining the phase estimation technique based on the Fourier Quantum Transform with the Grover iteration \citep{nielsen2002quantum}. Alternatively, it is also possible to apply the amplitude estimation to the problem of approximate counting \citep{brassard5055quantum}, or to use techniques inspired by Shor's celebrated quantum factorization algorithm and combines them with Grover's algorithm \citep{boyer1998tight}. Note that the number of solutions is known at this moment, but their search space positions are not known.

With the number $k$ of solutions determined, the states will be constructed according to Equation \ref{eq:redefinicao-base-quatro-estados-rotulo}. The quantum walk occurs in a complete graph, as shown in Figure \ref{fig:complete-graph-with-selfloops}, where the quantum walk will be performed in a four-dimensional space according to in Section \ref{sec:lackadaisical-quantum-walk-in-complete-graph}, reducing the search space.  Thus, the search space is represented by the superposition of vertices and edges already presented in Equation \ref{eq:redefinicao-base-quatro-estados-rotulo}.

Following the Algorithm \ref{alg:algoritmo-em-grafo-completo} in line $4$, the initial state preparation is performed considering the space size, the number of solutions, and the relations between vertices and edges, according to Equation \ref{eq:estado-inicial-grafo-completo-redefinido}.

After preparing the system, the quantum walk is performed. Line $5$ of the Algorithm \ref{alg:algoritmo-em-grafo-completo} defines the total number of steps $t$, according to Equation \ref{eq:qtd-passos-probabilidade-maxima}. At each step $j$, the evolution operator $U$ is applied to the quantum system $\ket{\Psi_{j-1}}$, where $j = 1,2,\dots,t$, as shown in line $6$.

Once the evolution is completed, a measurement in the basis $\ket{x}\ket{x \rightarrow y}$ is performed, as presented in line $8$. Thus, the states $\ket{a}\ket{a \rightarrow a'}$ or the states $\ket{a}\ket{a \rightarrow b}$, which both have the solutions, is obtained with high probability. It is worth noting that the vertex information where the walker stays defines the found solution at the time of the measurement. The direction where the walker points for the next quantum movement (defined by its edge) is only relevant to define the quantum walk evolution, not to define the solution state at the measurement time.

After recovering the states $\ket{a}\ket{a \rightarrow a'}$ or $\ket{a}\ket{a \rightarrow b}$ with high probability after the measurement, in the line $9$, the algorithm initializes the classical neural network weights as shown on the right side of Figure \ref{fig:Lattice}.

\begin{algorithm}
\SetAlgoLined
\Begin{
Set the parameters: $\Delta p$, $N$ and $l$\\
Quantum count execution\\
Preparation of the initial state\\
\For{$j \leftarrow 1$ \KwTo
$\frac{\pi}{\sqrt{2(2k+l-1)}}\sqrt{N} $}{
$\ket{\Psi_{j}} \leftarrow U\ket{\Psi_{(j - 1)}}$\\
}
Make the measurement \\
Initialize the weights of the Artificial Neural Network
}
\caption{\textsc{Training Algorithm.}}
\label{alg:algoritmo-em-grafo-completo}
\end{algorithm}
\section{Experiment Setup}
\label{sec:experiment-setup}

A simple classification problem was performed to assess the concept of the proposed algorithm.  An artificial neural network of MLP type (Multilayer Perceptron Type) was employed to solve the ``EXCLUSIVE-OR'' classification problem. The neural network was initialized with the weights generated by the procedure proposed in this work.

\subsection{Exclusive-OR Function}
\label{sec:exclusive-or-function}

The EXCLUSIVE OR (XOR) problem, a simple but nonlinearly separable problem, was used as the function to be learned by the neural network. The XOR function, also known as exclusive disjunction, is an operation on two binary values, $x_{0}$ and $x_{1}$, where if only one of these binary values is equal to $1$, then the function returns $1$, otherwise returns $0$ \citep{lecun2015deep}.

Consider the problem of classifying points in the unitary hypercube. The EXCLUSIVE OR can be understood as a particular case of this problem. In this case, it is sufficient to consider only the four vertices of the unit square corresponding to the points $\{(0,0),(0,1),(1,0),(1,1)\}$. Each set of patterns determine outputs that are called classes. The inputs $\{(0,0),(1,1)\}$ generate outputs $x_{0} \oplus x_{1} = 0$, where it will be called class $0$. The inputs $\{(0,1),(1,0)\}$ generate outputs $x_{0} \oplus x_{1} = 1$,  class $1$ \citep{haykin}. Therefore, it is a pattern classification problem that consists of associating an input pattern $(x_{0}, x_{1})$ with one of the previously defined classes $\{0,1\}$ \citep{da2017artificial}.

\subsection{Neural Network Architecture}
\label{sec:neural-network-architecture}

The EXCLUSIVE OR problem can be solved by a Multi-Layer Perceptron (MLP) neural network with three neurons, where two of them are in the hidden layer, and the other neuron is in the output layer. The neural network used in this work follows this architecture. 

\begin{figure}
    \centering
    \includegraphics[height=6cm]{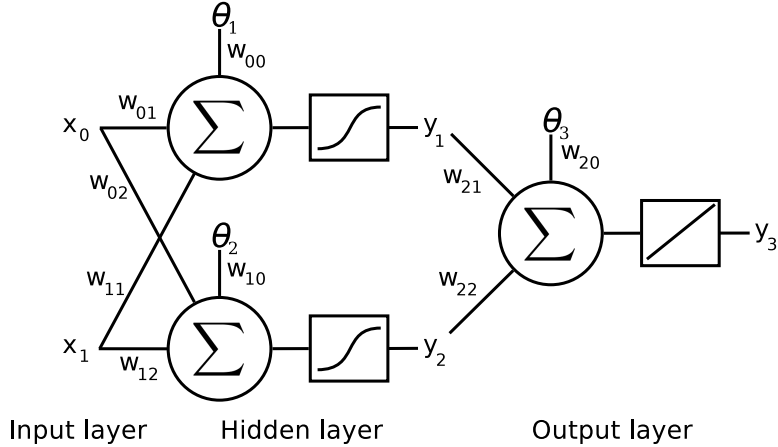}
    \caption{Configuration of a multilayer neural network. The letters $ \theta_1 $, $ \theta_2 $ and $ \theta_3 $ mean the bias and the letters $ y_1 $, $ y_2 $ and $ y_3 $ mean the outputs of neurons.}
    \label{fig:neural-network-section-4-2}
\end{figure}

The input layer has two values $\{\textrm{\fontfamily{phv}\selectfont x}_{0}, \textrm{\fontfamily{phv}\selectfont x}_{1}\}$ that are inputs data. The hidden layer has two neurons. The output layer has one neuron.  Each neuron has a bias. Thus, the neural network has nine synaptic weights. Six weights are in the hidden layer $\{\omega_{00}, \omega_{01}, \omega_{02}, \omega_{10}, \omega_{11}, \omega_{12}\}$ and three are in the output layer $\{\omega_{20}, \omega_{21}, \omega_{22}\}$ according to Figure \ref{fig:neural-network-section-4-2}. The sigmoid logistic, described by Equation \ref{eq:sigmoid-logistica}, is the activation function for all neurons of the hidden layer and the neuron of the output layer is linear,

\begin{equation}\label{eq:sigmoid-logistica}
    f(x) = \frac{L}{1 + e^{-\lambda(x-\zeta_{0})}}
\end{equation}
where $\lambda = 1$ is the declivity of the curve, $L = 1$ is the maximum value of the curve, and $\zeta_{0} = 0$ is the value of $x$ at the midpoint of the curve.

\subsection{Hardware and Software Setup}
\label{sec:hardware-software-setup}

The simulations were performed using the following hardware configurations. For simulate neural network training with classical backpropagation algorithm, it was used,

\begin{itemize}
    \item[$\bullet$] Operational System: Debian GNU/Linux 10 Buster;
    \item[$\bullet$] Memory: 4 GiB;
    \item[$\bullet$] Processor: Intel Core i3-5005U CPU @ 2.00 GHz x 4;
    \item[$\bullet$] OS type 64-bit;
    \item[$\bullet$] HD: 1 terabytes.
\end{itemize}
To simulate the procedure using a quantum walk in the complete graph, it was used

\begin{itemize}
    \item[$\bullet$] Operational system: Debian GNU/Linux Jessie 8.11;
    \item[$\bullet$] Memory: 16 GiB;
    \item[$\bullet$] Processor: Xeon Intel 5th gen CPU @ 3.6 GHZ x 8;
    \item[$\bullet$] OS type 64-bit;
    \item[$\bullet$] HD: 4 terabytes.
\end{itemize}

The programming languages used to write the algorithms were Python 3.6 with the open-source machine learning framework PyTorch\footnote{\url{https://pytorch.org/}}.

\subsection{Search Space}
\label{sec:Search space}

Theoretically, the quantum walk is performed in an infinite space. However, in practice, because of memory and hardware limitations, we have determined sub-spaces or windows. Once the window size is defined, the procedure performs the search within it. 

Geometrically, the windows employed here will always be hyper-cubes in the search space. For example, in the $2$-dimensional case, a window is a square with $N = z^2$ points, where $z$ is the number of points in a dimension. For the $d$-dimensional case, a window is a hyper-cube whit $N=z^d$ points, where there are $z$ points in each dimension.  

\begin{figure}
\centering
\includegraphics[height=6cm]{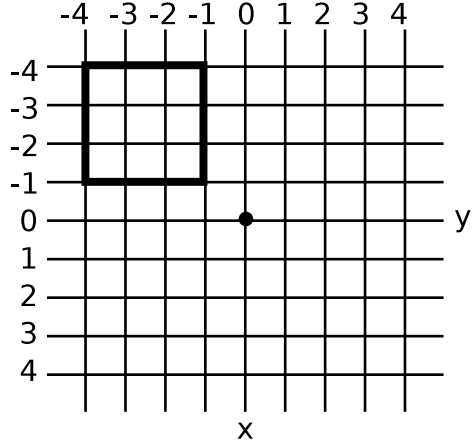}
\caption{Illustration of a search in an infinite two-dimensional grid using a window with a finite set of points.}
\label{fig:search-two-dimensional-infinite-with-windows}
\end{figure}

If there is no solution in the sub-region defined by the windows, the windows will be shifted. By applying offsets, the windows are moved by performing the search in the infinite search space regions. To illustrate, consider an infinite two-dimensional search space as illustrated in Figure \ref{fig:search-two-dimensional-infinite-with-windows}. These window shifts in infinite space are carried out until at least one solution is found in the current window. There are many forms to define the shift of the windows. However, the simplest way is to sum an offset of size $z$ for each dimension with respect to the current window, which was the approach used in this work.

\subsection{Representation of the Oracle}
\label{sec:representation-oracle}

\begin{figure}
\centering
\includegraphics[height=3cm]{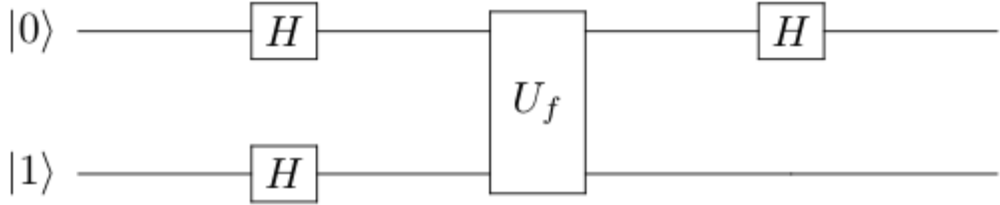}
\caption{The circuit that implements the Deutsch algorithm \citep{yanofsky2008quantum}.}
\label{fig:algoritmo-deutsch}
\end{figure}

An oracle is a structure capable of generating answers to binary questions. The circuit shown in Figure \ref{fig:algoritmo-deutsch} represents the implementation of the Deutsch algorithm. Conditional port $U_{f}$ implements the NOT-controlled port with control bit $f(x)$ and acts as $U_{f}:\ket{x,y} \rightarrow \ket{x,y\oplus f(x)}$.

Port $U_{f}$ is a black box with no explicit implementation, often called an oracle \citep{sasaki2013quantum}. In this way, a quantum oracle is a ``black box'' operator that, when applied to a system state, return if this specific state is a solution or is not a solution. An example of oracle implementation can be seen in the work of \citet{zheng2018quantum}, which uses Grover's oracle as a central part of the circuit that implements their proposal of quantum perceptron models \citep{kapoor2016quantum}.

In this way, it was necessary to create a representation of an oracle to perform the simulations. The solution employed uses a sparse matrix with the size of the search space. The positions filled with $1$ indicate the points that are a solution and $0$ otherwise. In practice, the oracle answers whether a given point in the walker space, after converted to synaptic weights, correctly classifies the input patterns or not. However, a formal conception of this quantum oracle for this purpose is beyond the scope of this work. Therefore, we considered it existing.

\subsection{Weight Generation}
\label{sec:weight-generation}

The quantum walk performs a search in quantum states that represent points in an integer space. At the end of the process, it is necessary to convert these points into synaptic weights values represented by real numbers. Thus, a real value was defined, $\Delta p$, where it will multiply each measured point at the end of the process. That is, the walker grid space is in $\Delta p$ units. The walker starts at the window center, and the walker will move by the positive and negative integer indices of each dimension of the grid.  

Therefore, when the proposed algorithm makes the measurement at the walker position in the grid $\ket{r_i}$ (where $i$ is the label for each point in the $w$-dimensional search space), the integer components of $r_i = (n_1, n_2, \dots, n_w)$ for each dimension are converted to the real values $\Delta p*n_j$ ($j=1,2,\dots,w$), generating the synaptic weights for the neural network. In this way, the search is performed only among the factors of $\Delta p$. Also, the value defined for $\Delta p$ establishes the search refinement level.

\subsection{Measurement Process}
\label{sec:measurement-process}

Through unitary transformation, closed quantum systems evolve. In order to be able to access information that is in the $\ket{\Psi}$ state, an observation must be made. In practice, measurements are made in laboratories using physical devices, such as lasers, magnets, scales, and stopwatches, but in theory, this measurement process is described mathematically \citep{portugal2013quantum}. According to the postulate of quantum mechanics, the probability of a state occurs is,

\begin{equation}
\label{eq:before-the-measurement}
    p(m) = \braket{\Psi \mid M^{\dagger}_{m}M_{m} \mid \Psi},
\end{equation}
where $M$ is a measurement operator, $\dagger$ is the symbol used to describe the conjugate transpose operation, and the index $m$ represents the results that can occur in the experiment after the measurement. The state $\ket{\Psi'}$ described in Equation \ref{eq:after-the-measurement} is the state of the system just after the measurement \citep{nielsen2002quantum}.

\begin{equation}
\label{eq:after-the-measurement}
    \ket{\Psi'} = \frac{M_{m}\ket{\Psi}}{\sqrt{\braket{\Psi \mid M^{\dagger}_{m}M_{m} \mid \Psi}}}
\end{equation}

As an example we are going to measure the state $\ket{\Psi}$ whose states are described in Equation \ref{eq:redefinicao-base-quatro-estados-rotulo}. Considering the system with four possible results we will define the operators $\{\ket{AA}\bra{AA}, \ket{AB}\bra{AB}, \ket{BA}\bra{BA}, \ket{BB}\bra{BB}\}$. So when applying Equations \ref{eq:before-the-measurement} and \ref{eq:after-the-measurement} we have,

\begin{equation*}
\label{eq:p-m}
    p(m) = \left | \lambda_{m} \right |^2
\end{equation*}

\begin{equation*}
\label{eq:psi-line}
    \ket{\Psi'} = \frac{\lambda_{m}}{\left | \lambda_{m} \right |}\ket{m}
\end{equation*}
where $m = AA,AB,BA,BB$.

In this way, the measurement is a probabilistic process, where theoretically, the probability of measuring a given state $\ket{m}$ will be $|\lambda_m|^2$. Thus, after the quantum walk evolution, it is possible to observe from the final quantum state the quantity $|\lambda_m|^2$ for each state component $\ket{m}$.

Statistically, as we can see in Section \ref{sec:experimental-results-and-discussions}, the chances of measuring the states that contain a solution were much more significant than measuring a state that did not contain a solution.

\subsection{Comparison against Backpropagation Algorithm}
\label{sec:comparison-backprop}

Since the search space contains solutions to the problem, the approach proposed in this paper can find a solution with high probability after the number of iterations defined in Equation \ref{eq:qtd-passos-probabilidade-maxima}. It implies that, in addition to ensuring a successful search, the number of iterations required to reach the artificial neural network training is known \textit{a priori}.

The classical backpropagation algorithm seeks to minimize an error function based on the gradient descent methodology. For this reason, it can stagnate in local minimums. The practical implication is the possibility that the algorithm runs indefinitely and yet does not generate correct outputs. These limitations of a gradient descent methodology are overcome with the proposed quantum algorithm.

Furthermore, many experiments were done with both approaches to compare the proposed algorithm with the traditional backpropagation algorithm. The criterion to measure the performance was the number of iterations needed to find an ANN weight set that solves the XOR problem. The number of epochs for the backpropagation and the number of iterations for the proposed algorithm were observed.

Comparing the number of epochs of the backpropagation algorithm with the number of iterations of the procedure developed in this work was applied to measure how much better one approach is than the other. The efficiency of the proposed procedure is the order of $O(\sqrt{N/k})$. However, it was not possible to compare the efficiencies in terms of running time between our procedure and the backpropagation algorithm.
\section{Experimental Results and Discussions}
\label{sec:experimental-results-and-discussions}

A set of experiments with classical backpropagation was done to define a comparative baseline. Employing the same ANN architecture $2-2-1$ (two inputs, two hidden neurons, and one output -- according to Section \ref{sec:neural-network-architecture}), five different learning rates  $\eta = \{0.5, 0.1, 0.01, 0.001, 0.0001\}$ were investigated, where $1200$ simulations were computed for each one. The maximum number of epochs of $150000$, the training stagnation, and the zero classification error were the stopping conditions employed. The network was considered in a stagnation training situation if its MSE error did not decrease by $1000$ consecutive epochs. The classification error is zero when the ANN can classify all the four $XOR$ inputs correctly. 

For these conditions, none of the classical backpropagation experiments reached a stagnation situation. The experiments with learning rates of $0.5$ and $0.1$ never reached the maximum number of epochs, always reaching the perfect classification. The summarization of this experimental behavior can be viewed in Table  \ref{tab:backpropagation-simulations}, where \emph{lr} indicates the learning rates, \emph{Epochs limit} is the number of times that the experiment reached the maximum number of epochs stop condition, and \emph{Successful} is the number of experiments that obtained zero classification error.

\begin{table}
\centering
\caption{Classical Backpropagation Simulations. The column lr presents the investigated learning rate, Epochs limit shows the experiment number that reached the maximum number of epochs, and Successful the number of experiments with zero classification error.}
\label{tab:backpropagation-simulations}
\begin{tabular}{lcc}
\hline
lr &  Epochs limit & Successful
\\ \hline
0.5000 & None & 1200 \\
0.1000 & None & 1200 \\
0.0100 & 452 & 748 \\
0.0010 & 467 & 733 \\
0.0001 & 726 & 474 \\
\hline
\end{tabular}
\end{table}

Table \ref{tab:descriptive-backpropagation} presents the descriptive statistics for the baseline experiments. For each learning rate, it is presented the minimum, mean, maximum value of the Epochs number, and its standard deviation (Std.).

Table \ref{tab:experiments-procedure} presents the results of the simulations performed using the proposed computational procedure described in Algorithm \ref{alg:algoritmo-em-grafo-completo}. Here, $k$ represents the number of solutions, $N$ the number of vertices, $l$ the number of self-loops (for all experiments $l = 1$), and $t$ the number of iterations. $\Delta p = 0.5$ was utilized for all experiments. Here, it was investigated three configurations of $N \,(2^9 = 512;\, 4^9 = 262144;\, 8^9 = 134217728)$ in five experiments. Two experiments with $N=512$, 2 points per dimension. Two experiments with $N=262144$, 4 points per dimension, and one experiment with $N=134217728$, 8 points per dimension. The search spaces are sub-regions (windows) of an infinite $\omega$-dimensional grid, as presented in Section \ref{sec:Search space}. However, the initialization of these sub-regions can occur in different positions of the infinite grid. If there are no solutions in the current sub-region, the sub-region is moved to another region of the infinite grid. All simulations had their sub-region initiated randomly around the origin. The shift of the sub-region occurs until a region with at least one solution is found. With a few interactions, the sub-space converged to a region with solutions in all cases studied here.

Coincidentally, although the experiments 1 and 2 used distinct search windows, the proposed procedure converged to sub-spaces with 12 solutions (see Table~\ref{tab:experiments-procedure}). As $N = 512$ and $k = 12$ for those experiments, the iteration number $t$ is also equal, being $t=11$, which was theoretically defined to $10.26$ by Equation~\ref{eq:qtd-passos-probabilidade-maxima}. For the simulations 3 and 4, the proposed procedure converged to sub-spaces with 17 and 20 solutions, respectively. Thus, the iteration number $t$ is different for those simulations, being $196$ and $180$, theoretically defined to $195.83$ and $179.83$, respectively. For the simulation 5, the procedure converged to a sub-space with more than eighty thousand solutions ($80295$), which implied in $t = 65$ iterations (theoretical number of iterations of $64.22$).

The procedure developed using the quantum search algorithm, even in spaces with a high number of vertices, was able to amplify the amplitudes in a relatively low number of iterations compared with the results for the classical backpropagation procedure. In the best case, the maximum iteration number for the quantum walk algorithm was $11$, while the mean number of epochs obtained by the backpropagation algorithm (in the best case) was $33$. 

It is also possible to see in Table \ref{tab:descriptive-backpropagation} that the minimum number of epochs of backpropagation is less than the number of the interactions of the quantum walk procedure, see Table \ref{tab:experiments-procedure}. Although it appears that the classical backpropagation algorithm has an advantage over the quantum walk procedure, it is essential to observe that the number of interactions $t$ is known before start the search in the quantum walk algorithm, but it is not for the classical backpropagation. Thus, the critical measure to characterize the practical cost expectation is the mean number of epochs for the classical backpropagation. Prior knowledge of the amount of interaction required for the algorithm to converge is a great advantage of the quantum walk algorithm.

\begin{table}
\centering
\caption{
Descriptive statistics. Parameters of the backpropagation algorithm experiment for the number of epochs. lr means the learning rate. Std. means Standard Deviation.
}
\label{tab:descriptive-backpropagation}
\begin{tabular}{p{0.15\linewidth}p{0.15\linewidth}p{0.15\linewidth}p{0.15\linewidth}p{0.15\linewidth}}
\hline
\multicolumn{5}{c}{\large Epoch Statistics}\\
\hline
lr & Minimum & Mean & Maximum & Std. \\ \hline
       0.5000 &    1 &         33.60 &     319 &    35.68 \\ 
       0.1000 &    3 &        433.84 &    3279 &    463.78 \\
       0.0100 &    2 &       5277.48 &  132199 &    17927.67 \\
       0.0010 &    9 &      12949.18 &  148256 &    22451.79 \\
       0.0001 &  295 &      46987.00 &  149644 &    36780.22 \\
       \hline
\end{tabular}
\end{table}

Therefore, observing the mean result of epochs required for network training by the backpropagation (Table~ \ref{tab:descriptive-backpropagation}) and the values of $t$ in Table \ref{tab:experiments-procedure}, the computational proposed procedure is more efficient on average. Depending on the size of the search space of the proposed algorithm, the network training using the backpropagation algorithm, in some cases, performs a smaller number of iterations. However, on average, the iterations number of the proposed algorithm is decidedly smaller than the backpropagation procedure. As seen in the backpropagation simulations, the random initialization of weights is one factor that influences the result of time convergence for the network. However, because of the impossibility of determination for an excellent region to initialize the weights, the mean value and standard deviation are the information statistically relevant for backpropagation experiments, where low performance and high variation in training epochs number results are obtained when compared with the proposed quantum algorithm.

\begin{table}
\centering
\caption{Weight search experiment by the quantum walk procedure. K means the number of solutions. N is the total number of vertices. The number of iterations is represented for letter t. For all experiments, the number of self-loops was $l=1$.}
\label{tab:experiments-procedure}
\begin{tabular}{cllcc}
\hline
\multirow{2}{*}{Experiment} & \multirow{2}{*}{k} &  \multirow{2}{*}{N} & \multicolumn{2}{c}{t} \\ \cline{4-5}
  & & & Theoretical & Simulated \\\hline
1 & 12 & 512 & 10.26 & 11 \\
2 & 12 & 512 & 10.26 & 11 \\
3 & 17 & 262144 & 195.83 & 196 \\
4 & 20 & 262144 & 179.83 & 180 \\
5 & 80295  & 134217728 & 64.22& 65 \\ \hline
\end{tabular}
\end{table}

\begin{table}
\centering
\normalsize
\caption{Measurement experiment --- the percentage measure observed for each quantum. The states $\ket{AA}$ and $\ket{AB}$ are the solutions.}
\label{tab:measurements}
\begin{tabular}{ccccc}
\hline
Experiment & \multicolumn{1}{c}{$\ket{AA}$} & \multicolumn{1}{c}{$\ket{AB}$} & $\ket{BA}$ & \multicolumn{1}{c}{$\ket{BB}$} \\ \hline
1 & 95.48\% & 3.07\% & 1.40\% & 0.05\% \\
2 & 95.03\% & 3.67\% & 1.26\% & 0.04\% \\
\textbf{3} & \textbf{100.0\%} & 0.00\% & 0.00\% & 0.00\% \\
\textbf{4} & \textbf{99.99\%} & 0.00\% & 0.01\% & 0.00\% \\
\textbf{5} & \textbf{99.88\%} & \textbf{0.10\%} & 0.02\% & 0.00\% \\ \hline
\end{tabular}
\end{table}

Nevertheless, the proposed algorithm is quantum. Table~\ref{tab:measurements} shows the measurement probabilities for the five simulations for neural network training. It was considered both states $\ket{AA}$ and $\ket{AB}$ are solutions, given that for these states, the walker is in a vertex $a$. At least $98.55\%$ of the measurements find a solution (experiment $1$), reaching $100\%$ for the experiment $3$. At mean, the proposed quantum procedure find a solution in $99.44\%$ of the measurements.
\section{Conclusion}
\label{sec:conclusion}

Training based on the backpropagation algorithm (or descendant gradient algorithms) may fall to local minimums. Many factors can influence this result and carry training for many epochs until the network converges to a solution or even stagnates. Therefore, it is impossible to say whether the training will stop or that good accuracy can be guaranteed in a prior way.

The non-stagnating in local minimums and the knowing in advance the number of iterations required to obtain a solution are some advantages of the proposed procedure. It is not possible to guarantee that the solution obtained is optimal because the probability is equal between all weight set solutions within the state $\ket{AA}$ and the state $\ket{AB}$. However, with a high probability, there will have a valid solution at the end of the procedure. So, the procedure proposed here guarantees the neural network training with high probability.  However, the proposed methodology is a theoretical proposal. In practice, the proposed algorithm needs a quantum computer, which does not exist yet.

Another critical point is the oracle. In a quantum system, an oracle is an operator that can answer if a given state is or is not marked, \textit{i.e.}, if a given state is or is not a solution. For the proposed methodology, an oracle would be an operator capable of determining whether a given state would train a network or not.
Here, an oracle was simulated by a simple matrix of zeros and ones. Each position in the oracle matrix is a possible state in the quantum system. If the value is $1$, then the state is a solution. Otherwise, the state is not a solution. The creation of a real quantum oracle operator is a future research.

As seen before, the execution time of the proposed algorithm is of the order of $O (\sqrt{N/k})$, \textit{i.e.}, there a quadratic gain when compared with the classical analog. In this case, the search problem grows exponentially, the window's search space is $\mathbf{O}(P^{\omega})$, where $P = \sqrt(N)$, $N$ is the number of points in the search window (a squared window), and $\omega$ is the number of ANN weights. The number of qubits required to represent $\omega$ synaptic weights is equal to $\omega * \log_2(N)$. The proof of correctness of the quantum walk algorithm can be found in \citep{feng2007proof}.

Finally, classical simulations showed a significant gain in training an artificial neural network using the procedure that applies a quantum walk to find the set of weights compared with the use of the backpropagation algorithm.

\section*{Acknowledgments}
\label{sec:acknowledgments}

Acknowledgments to the Science and Technology Support Foundation of Pernambuco (FACEPE) Brazil, Brazilian National Council for Scientific and Technological Development (CNPq), and Coordena\c{c}\~{a}o de Aperfei\c{c}oamento de Pessoal de N\'{i}vel Superior - Brasil (CAPES) - Finance Code 001 by their financial support to the development of this research.

\bibliographystyle{unsrtnat}






\end{document}